\documentclass[onecolumn]{webofc}
\usepackage[varg]{txfonts}   
\usepackage{graphicx}
\usepackage[justification=centering]{caption}
\usepackage[justification=justified,singlelinecheck=false]{caption}
\begin{document}

\title{Stellar \textit{s}-process neutron capture cross sections on $^{A}$Se and $^{A}$Ce  }

\author
{\firstname{R. N.} \lastname{Sahoo}
\inst{1}
\and
\firstname{M.} \lastname{Tessler}\inst{1}
\inst{2}
\fnsep
\thanks{\email{moshe.tessler@mail.huji.ac.il}} 
\and
\firstname{S.} \lastname{Halfon}\inst{2}
\and
\firstname{Y.} \lastname{Kashiv}\inst{3}
\and
\firstname{D.} \lastname{Kijel}\inst{2}
\and
\firstname{A.} \lastname{Kreisel}\inst{2}
\and
\firstname{M.} \lastname{Paul}
\inst{1}
\fnsep
\and
\firstname{A.} \lastname{Shor}\inst{2} 
\and
\firstname{L.} \lastname{Weissman}\inst{2}
}

\institute{
The Hebrew University of Jerusalem, Jerusalem, Israel 91904
\and
Soreq Nuclear Research Center, Yavne, Israel 81800
}

\abstract{%
We report on experiments at the Soreq Applied Research Accelerator Facility - Liquid-Lithium Target (SARAF-LiLiT) laboratory dedicated to the study of s-process neutron capture reactions. The kW-power proton beam at 1.92 MeV (1-2 mA) from SARAF Phase I yields high-intensity 30 keV
quasi-Maxwellian neutrons (3-5×10$^{10}$ n/s). The high neutron intensity enables Maxwellian averaged cross sections (MACS) measurements of samples with short-lived decay products. Neutron capture reactions on $^{\textit {nat}}$Se and $^{\textit {nat}}$Ce were investigated by activation in the LiLiT neutron beam and $\gamma$-spectrometry measurements of their decay products.}
\maketitle
\vspace{-0.7 cm}
\section{Introduction}
\label{intro}
Except for the lightest nuclei (H, He, Li) that were created during the nucleosynthesis era of the Big Bang, the nuclei of the elements are produced in stars. Neutrons generated in-situ, and in particular their captures on nuclear seeds, are responsible for the production of the
large majority of the heavier nuclides (A $\ge$ 60) \cite{B2FH}. Neutron-induced reactions remain at the forefront of experimental investigations for the understanding of stellar nucleosynthesis and chemical evolution of the Galaxy in the region of medium- and heavy-mass nuclides \cite{KAP11}.
One of the neutron-induced nucleosynthesis paths, namely the so-called slow neutron capture process (s-process) consists of a succession of neutron captures, each occurring over a mean time longer than or commensurate with the typical mean life of $\beta$-decaying nuclides close to
the valley of stability. The s-process is believed to occur in stars during their He core and shell burning stages, where neutrons are available from ($\alpha$, n) reactions. In contrast, the rapid neutron capture process (r-process) consists of the sequential captures of a large number ($\approx$ 20) of neutrons on a fast time scale, much shorter than $\beta$-decaying times, occurring in an astrophysical site with an extreme neutron density ($\approx$ 10$^{20}$ - 10$^{22}$ cm$^{-3}$) \cite{KAP11}.

We report on measurements of the cross section of neutron capture reactions $^{\text {74,80,82}}$Se(\(n, \gamma\)) and $^{\text {136,138,140,142}}$Ce(\(n, \gamma\)) relevant, respectively, to the weak and main s-processes. The $^{A}$Se data pursues our recent study of the $^{69,71}$Ga stellar (\(n, \gamma\)) reactions in the weak s-process regime \cite{TES22}. The proton rich isotope $^{74}$Se is a \textit{p} nuclide, shielded from the \textit{s} and \textit{r} processes by stable nuclei in the region. The disentanglement of the different
heavy-nuclide synthesis modes (\textit{s-, r-} and \textit{p}-processes) requires reliable and precise stellar neutron-capture cross sections. Such is the case also for the Ce isotopes in the main \textit{s}-process \cite{KAP96}. In particular, $^{140}$Ce is found to be one of the most important nuclides in the network of \textit{s}-process reactions, affecting the abundances of a large number of isotopes \cite{KOL16}. The experiments were performed by the activation technique using a high-intensity (3-5×10$^{10}$ n/s) quasi-Maxwellian neutron beam that mimics conditions of stellar \textit{s}-process nucleosynthesis. The neutron field was produced by a mA proton beam at E$_p$=1925 keV (beam power of 2–3 kW) as part of our experiment campaign at the Phase I of Soreq Applied Research Accelerator Facility (SARAF) \cite{MAR18}, bombarding the Liquid-Lithium Target (LiLiT) \cite{HAL14, PAU19a}. The cross sections were measured by counting the activity of neutron capture products via $\gamma$-spectrometry with a high-purity germanium (HPGe) detector.

\section{The Soreq Applied Research Accelerator Facility (SARAF) and the Liquid-Lithium Target (LiLiT): activation experiments}

\begin{figure}[h!]
\centering
\includegraphics[scale=0.380, trim=0.0cm 0.1cm 0cm 0.0cm]{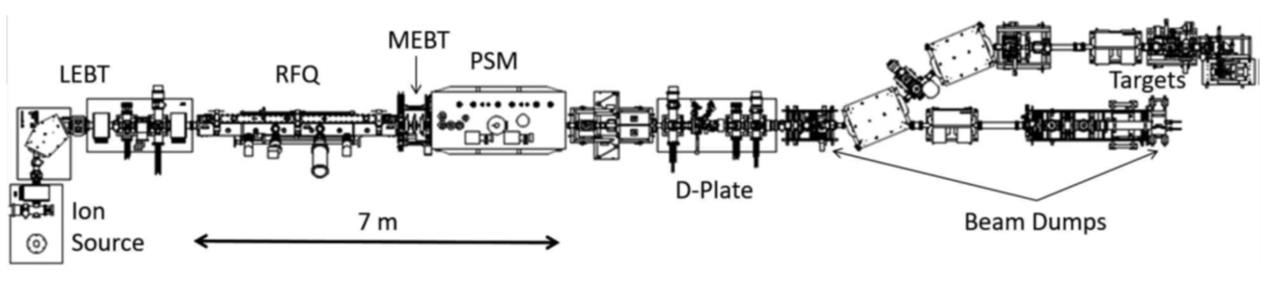}
\caption{Layout of SARAF Phase I; see text for acronyms} 
\label{fig1} 
\vspace{-0.5 cm}
\end{figure}

The SARAF accelerator \cite{MAR18} is a superconducting linear RF accelerator designed to produce a beam of light ions (m/q $\leq 2$) with a continuous-wave (CW) beam intensity up to 5 mA. The layout of Phase I is shown in Fig. \ref{fig1}. 

Light-ion H$^{+}$ or D$^{+}$ are extracted at 20 keV/u from an Electron Cyclotron Resonance (ECR) ion source, mass analyzed, and injected via a Low-Energy Beam Transport (LEBT) line to a 176 MHz four-rod Radio-Frequency Quadrupole (RFQ) with a final energy of 1.5 MeV/u. A short Medium-Energy Beam (MEBT) line transports the beam to a Prototype Superconducting Module (PSM) housing six Half-Wave Resonators (HWR), followed by the High Energy Beam (HEBT) line towards the LiLiT target at Station 1 (Fig. \ref{fig1}) via two 45$^{\circ}$ bending magnets.

\begin{figure}[h!]
\centering
\includegraphics[scale=0.320, trim=0.0cm 0.0cm 0.0cm 0.5cm]{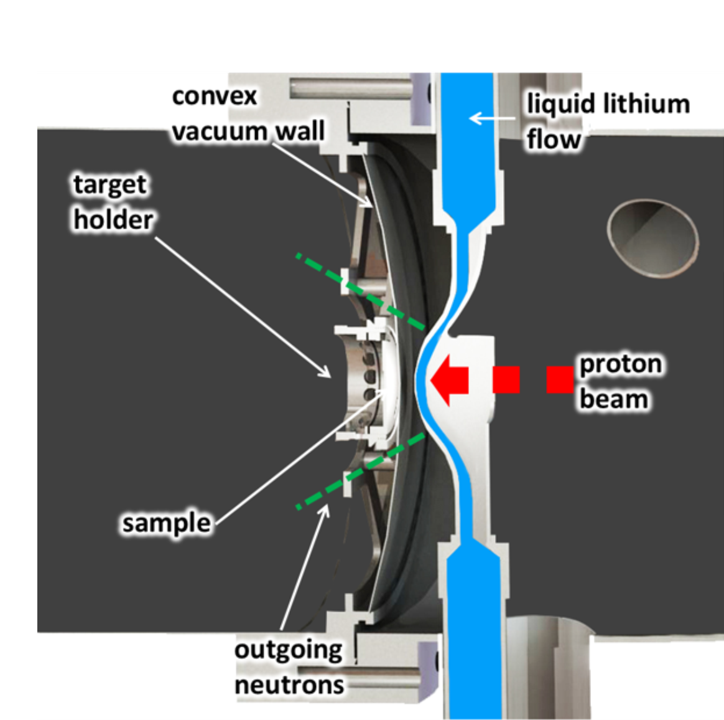}
\includegraphics[scale=0.320, trim=0.0cm 0.0cm 0.0cm 0.5cm]{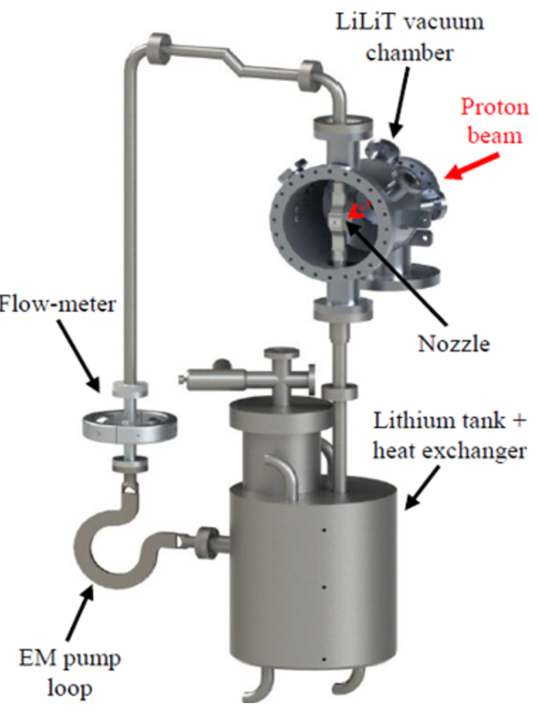}
\caption{ {(right) Schematic drawing of the Liquid-Lithium Target (LiLiT). (left) Diagram of the Liquid-Lithium Target (LiLiT) and activation target assembly. The (1-2 mA, $\approx$ 9 mm full width) proton beam (dashed red arrow) impinges directly on the windowless liquid-lithium film. The light blue shows the liquid-lithium circulating flow (see \cite{HAL14} for details). The activation samples are mounted at the center of a ring target holder made of Al and positioned in the outgoing neutron cone (green dashed lines) at a distance of $\approx$ 6 mm from the liquid-lithium film surface. The targets are in a vacuum chamber separated from the LiLiT chamber by a 0.5 mm stainless steel concave vacuum wall.}} 
\label{fig2} 
\vspace{-0.7 cm}
\end{figure}

The Liquid-Lithium Target (LiLiT) \cite{HAL14} was specifically designed and built to dissipate the high power of the CW proton beam from SARAF (up to $\approx$ 5 kW) and produce quasi- Maxwellian neutrons for nuclear astrophysics experiments (Fig. \ref{fig2}).

Liquid lithium is forced-flown from the top into a 15 mm wide, 1.8 mm thick film supported by a thin stainless steel wall (concave as viewed in the proton beam direction). The proton beam impinges directly (windowless) onto the lithium surface, emitting neutrons mainly in the forward direction. The liquid lithium acts both as a (thick) target producing neutrons near the surface ($\approx$ 4 $\mu$m deep) and as a beam dump. The high power ($\approx$ 0.7 MW/cm$^{3}$) at the Bragg peak depth is mechanically transported by the fast flow (2-5 m/s) to a reservoir and heat exchanger.

An activation measurement consists of irradiation of a secondary target in the neutron field generated by the $^{7}$Li(p, n) reaction in LiLiT. An experimental cross-section for the neutron-induced reaction is extracted from the offline decay counting of the activated target or by atom counting of the residual nuclide. The target must be positioned as close as possible to the Li film to maximize the neutron dose. The experimental setup is illustrated in Fig. \ref{fig2}. 

The secondary target is located in a so-called experimental chamber separated from the LiLiT chamber by a curved vacuum wall. The proton beam (full width of $\approx$ 9 mm) impinges on the free surface of the liquid lithium film, resulting in outgoing forward-directed neutrons. The wall curvature allows the secondary target within its holder to be at a distance of 6$\pm$1 mm from the liquid-Li surface. The secondary target is sandwiched within the holder between two Au foils serving as neutron fluence monitors. The setup and simulations were carefully benchmarked in~\cite{PAU19a,TES15} through measurements of the $^{94,96}$Zr(\(n, \gamma\))$^{95,97}$Zr cross sections, where a method of extrapolation of experimental cross sections to Maxwellian averaged cross sections (MACS) is described.

\section{$^{A}$Se and $^{A}$Ce measurements}

\begin{figure}[h!]
\centering
\includegraphics[scale=0.43, trim=0.0cm -0.3cm 0.0cm 0.5cm]{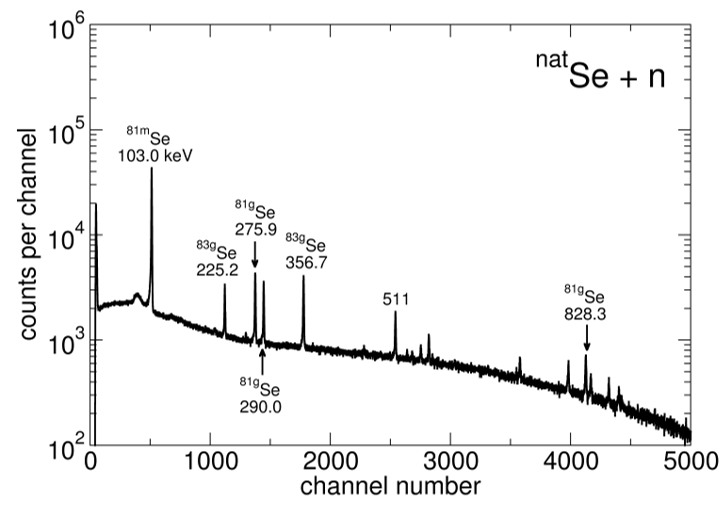}
\includegraphics[scale=0.20, trim=0.0cm 0.0cm 0.0cm 0.0cm]{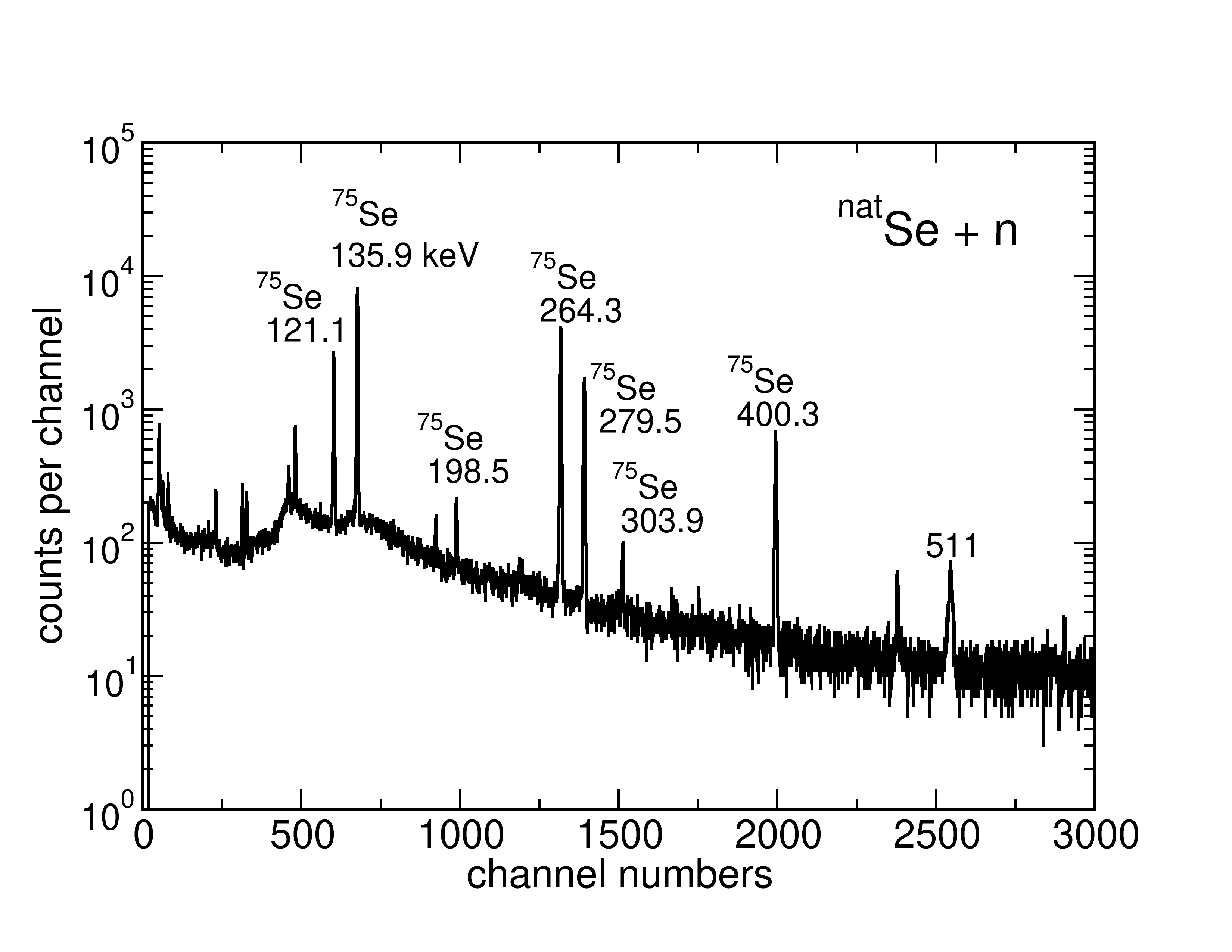}
\includegraphics[scale=0.20, trim=0.5cm 0.0cm 0.5cm 0.0cm]{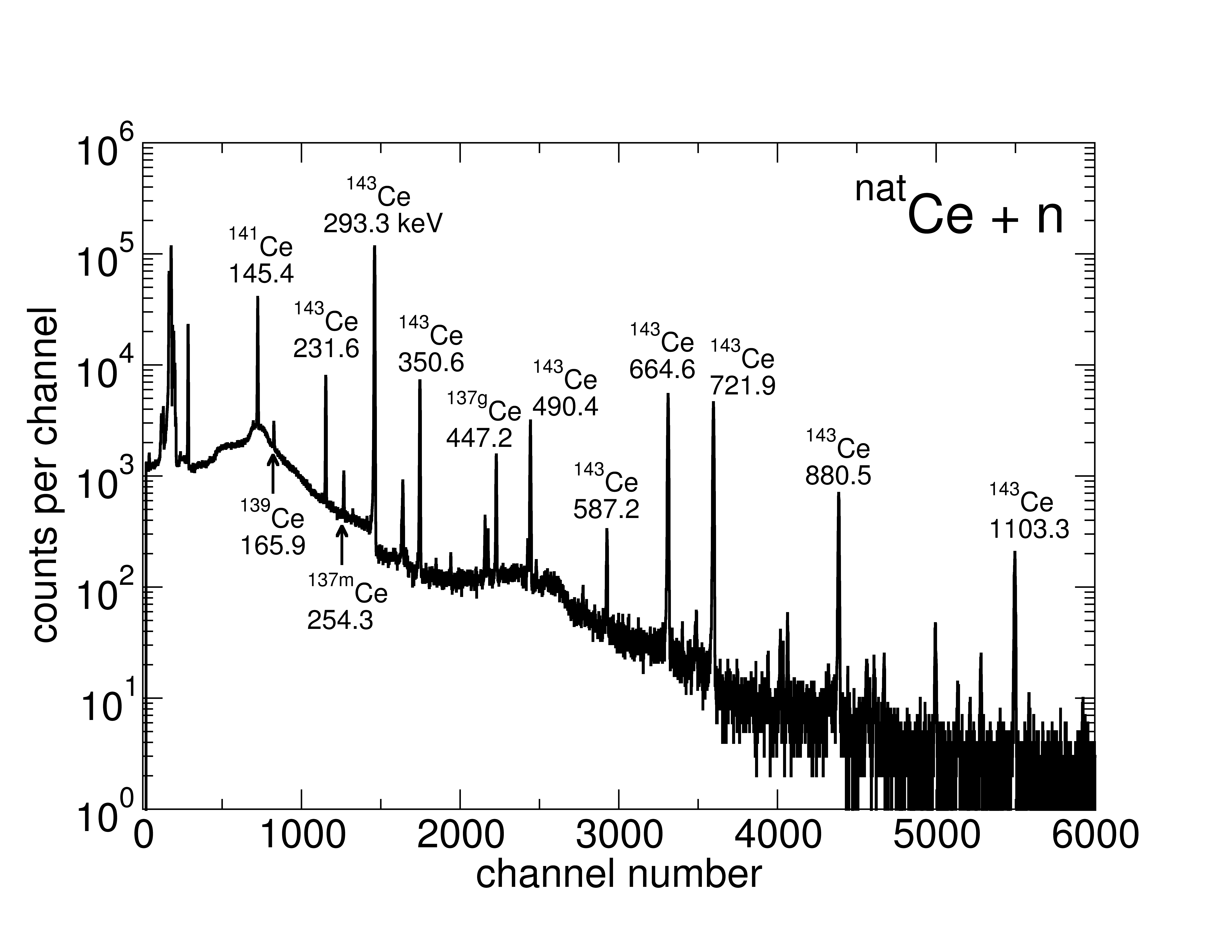}
\caption{Gamma spectra accumulated with a shielded High-Purity Ge detector for targets of 
\textsuperscript{nat}Se (top) and \textsuperscript{nat}Ce (bottom) activated in the neutron spectrum from LiLiT. The identified lines of the respective (\(n, \gamma\)) reaction products are indicated.  The top left spectrum is for the short-lived \textsuperscript{81,83}Se isotopes, and the top right spectrum is for the longer-lived \textsuperscript{75}Se isotope.} 
\label{fig3}
\vspace{-0.7 cm}
\end{figure}

$^{\textit {nat}}$Se and $^{\textit {nat}}$Ce targets were irradiated at SARAF-LiLiT for MACS measurements. After the irradiation, the induced activities are measured with a High-Purity Ge (HPGe) detector for each sample separately. Gamma spectra from the activated targets showing the major activation lines are displayed in Fig. \ref{fig3}.

When using a thick liquid-lithium target with a high-intensity proton beam, there is an ample production of high-energy $\gamma$ rays (17.6 MeV and 14.6 MeV) from the radiative capture $^{7}$Li(p, $\gamma$)$^{8}$Be reaction. In cases where there are a stable A-1 isotope, an unstable A isotope and a stable A+1 isotope, the nuclide A can be produced through the A--1(n,~$\gamma$) or A+1($\gamma$,~n) reactions.

\begin{figure}[h!]
\centering
\includegraphics[scale=0.20, trim=0.0cm 0.0cm 0.0cm 0.0cm]{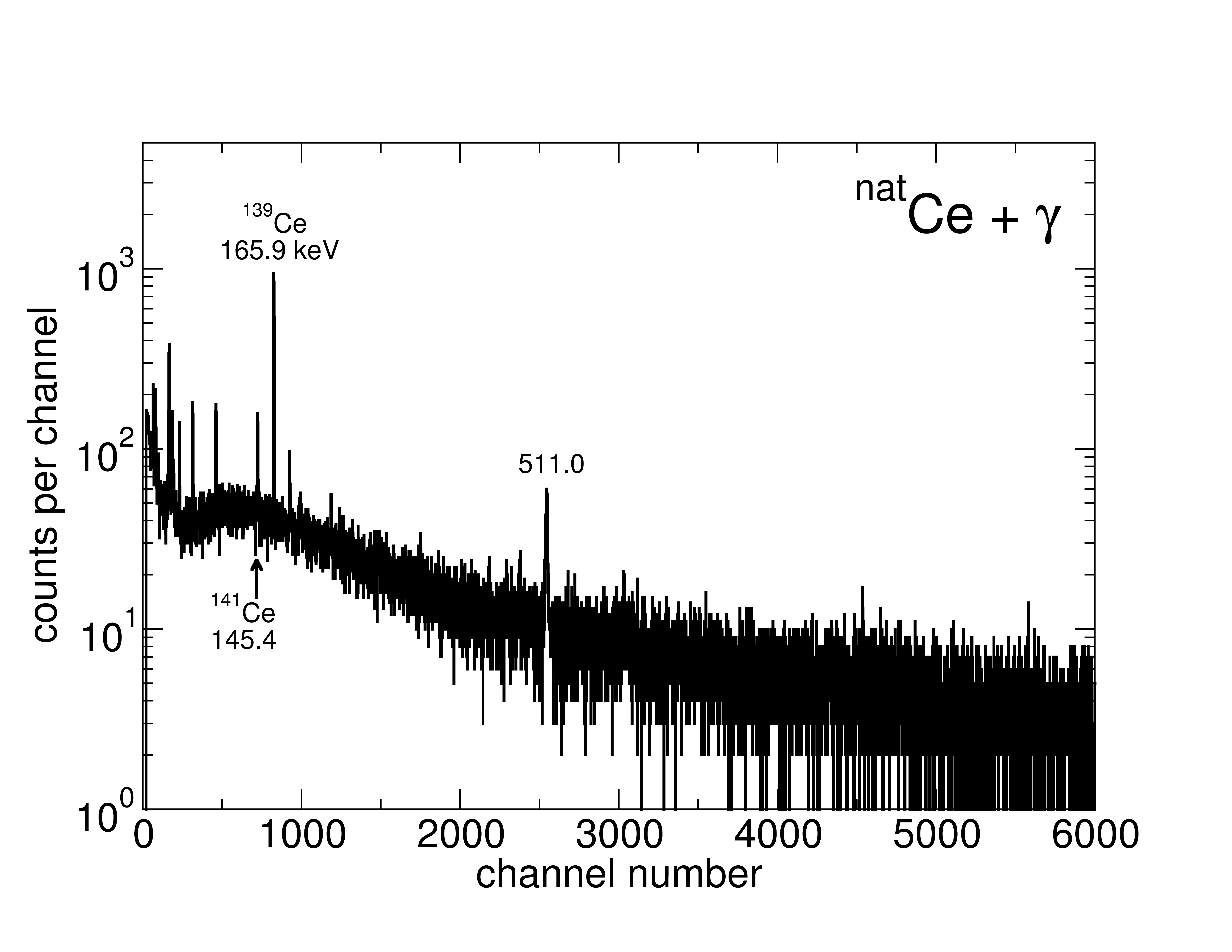}
\includegraphics[scale=0.44, trim=0.0cm 0.0cm 0.0cm 0.0cm]{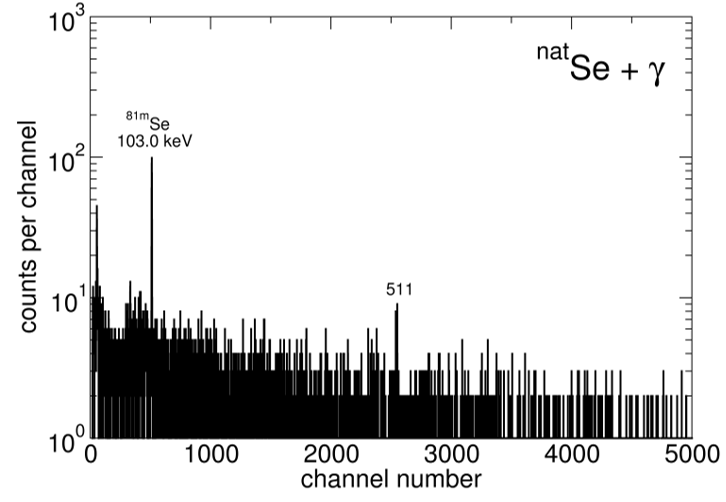}
\caption{Gamma spectra accumulated with a shielded High-Purity Ge detector for targets of \textsuperscript{nat}Ce (left) and \textsuperscript{nat}Se (right), activated below the neutron production threshold,  resulting in only gamma rays and no neutrons. 
The \textsuperscript{139,141}Ce and \textsuperscript{81\textit{m}}Se counts are then subtracted to obtain only the (\(n, \gamma\)) cross sections.} 
\label{fig4}
\vspace{-0.7 cm}
\end{figure}

In order to measure the contribution of the A+1($\gamma$,~n) reaction to the production of A, an irradiation of protons on LiLiT is conducted below the neutron production threshold. This results in the irradiation of the sample with $\gamma$-rays, but without neutrons. The ($\gamma$,\textit{n}) counts are then subtracted to obtain only the (\(n, \gamma\)) cross sections. The thick-target intensity of the 17.6 MeV and 14.6 MeV $\gamma$ rays at the SARAF-LiLiT setup from the $^{7}$Li(p, $\gamma$)$^{8}$Be reaction is estimated to be $\approx 7 \times 10^{8}$ $\gamma$ s$^{-1}$ per mA of proton beam at our relevant proton energy \cite{PAU19a,TES15,TES21}. Gamma spectra from the $\gamma$ activated targets are displayed in Fig. \ref{fig4}.

\begin{table}[ht]
\centering
\caption{Preliminary values (in mb) for MACS at 30 keV for the Se and Ce isotopes.}
\label{tabMACS}
\begin{minipage}{0.8\textwidth} 
\centering
\begin{tabular}{lcc}
\hline
{Nuclide} & {Recommended, } & {This work } \\
               & KADoNiS-1.0 \cite{KADV1} & {(preliminary)} \\
\hline
$^{74}$Se & 302(17) & 306(11) \\
$^{80}$Se $\rightarrow$ $^{81\text{m}}$Se & 8.6(9) & 9.3(4) \\
$^{82}$Se $\rightarrow$ $^{83\text{g}}$Se & 2.0(2) & 2.1(2) \\
$^{136}$Ce & 328(21) & 340(20) \\
$^{136}$Ce $\rightarrow$~$^{137\text{m}}$Ce & 28.2(16) & 28.9(14) \\
$^{138}$Ce & 179(5) & 201(11) \\
$^{140}$Ce & 11.73(44) & 8.5(5) \\
$^{142}$Ce & 29.9(10) & 26.4(7) \\
\hline
\end{tabular}
\end{minipage}
\end{table}

From the isotope activity relative to the Au activity (using the $^{197}$Au(n,~$\gamma$) cross section\cite{MAS10,LED11}), the experimental cross section is obtained. The Maxwellian Average Cross Section (MACS) at a given thermal energy $kT$ is then obtained by applying a correction to the experimental cross section (see~\cite{PAU19a,TES15,TES21} for details).

Preliminary results for the MACS are generally in good agreement with published values; Table \ref{tabMACS} lists preliminary results. Further analysis is in progress to reach final values that will be published when ready.
 
\section{Acknowledgments}
This work was supported by the Pazy Foundation (Israel). M.P. acknowledges support by the European Union (ChETEC-INFRA, project no. 101008324) and the German-Israeli Foundation (GIF), Grant Nr. 3013004592.


\end{document}